\documentclass[aps,prl,twocolumn,superscriptaddress]{revtex4-2}

\usepackage{makecell} % Para forzar saltos de línea en las celdas
\usepackage{microtype}%Afinar
\usepackage{graphicx}% Include figure files
\usepackage{dcolumn}% Align table columns on decimal point
\usepackage{bm}% bold math
\usepackage{caption} %Para colocar el caption de las imágenes
\usepackage{mathtools} % Para usar herramientas mathematics
\usepackage{tensor} % Para usar herramientas tensoriales
\usepackage[hidelinks]{hyperref} % Para crear enlaces
\usepackage{amssymb}%Símbolos mathematics
\usepackage{mathtools}%Más simbolos
\usepackage{mathrsfs}%Más símbolos
\usepackage{amsmath} % Para usar la matriz y los símbolos de derivadas parciales
\usepackage{xcolor}%Paquete para el color
\usepackage{array} %Para las tablas
\usepackage{booktabs}%Otra para tablas
\usepackage{subcaption}%Subreferencias
\usepackage{enumitem}
\usepackage{soul}%Resaltar
\usepackage{amsthm}
\usepackage[toc,page]{appendix}

\begin{document}

\title{Exact rotating dilatonic branch in ModMax electrodynamics without Maxwell analogue}

\author{Leonel Bixano}
    \email{Contact author: leonel.delacruz@cinvestav.mx}
\author{Tonatiuh Matos}%
 \email{Contact author: tonatiuh.matos@cinvestav.mx}
\affiliation{Departamento de F\'{\i}sica, Centro de Investigaci\'on y de Estudios Avanzados del Instituto Politécnico Nacional, Av. Instituto Politécnico Nacional 2508, San Pedro Zacatenco, M\'exico 07360, CDMX.
}%

%\affiliation{Departamento de F\'{\i}sica, Centro de Investigaci\'on y de Estudios Avanzados del IPN, Av. I.P.N. 2508, San Pedro Zacatenco, M\'exico 07360, CDMX.}
\date{\today}

\begin{abstract}
We present a novel class of rotating dilatonic solutions within the framework of Einstein-ModMax-type gravity. The configuration belongs to the nonlinear sector characterized by $\mathcal F/\mathcal G=\mathrm{const}$ and carries nontrivial electric and magnetic potentials, with both $A_t$ and $A_\varphi$ turned on, together with a nontrivial gravitomagnetic structure. We show that this solution does not admit continuation to the Maxwell framework of our parametrization, so it is intrinsically tied to the nonlinear ModMax regime. It includes both a NUT geometry and a NUT-free asymptotically flat limit, and it is valid for a broad class of dilatonic couplings, including the low-energy string and Kaluza-Klein cases. Moreover, in the prolate sector we identify a genuine black-hole regime in which the exterior region satisfies the null energy condition while the curvature singularity remains hidden behind the event horizon. These results provide an exact rotating dilatonic ModMax configuration with no Maxwell analog and a physically well-behaved exterior black-hole sector.
\end{abstract}

\maketitle

%%%%%%%%%%%%%%%%%%%%%%%%%%%%%%%%%%%%%%%%%%%%%%%%%%%%%%%%%%%%%%%%%%%%%%%%%%%%%%%%%%%%%%%%%%%%%%%%%%%%%%%%%%%%%%%%%%%%%%%%%%%%%%%%%%%%
%\section{Introduction}
%%%%%%%%%%%%%%%%%%%%%%%%%%%%%%%%%%%%%%%%%%%%%%%%%%%%%%%%%%%%%%%%%%%%%%%%%%%%%%%%%%%%%%%%%%%%%%%%%%%%%%%%%%%%%%%%%%%%%%%%%%%%%%%%%%%%
Introduction. Nonlinear electrodynamics offers a natural setting to investigate strong-field deviations from Maxwell theory in systems with gravity. Within this class of theories, ModMax is especially distinguished: it is the only one-parameter nonlinear generalization of Maxwell electrodynamics in four dimensions that maintains both conformal symmetry and continuous electric-magnetic duality. A key issue is thus whether its genuinely nonlinear regime admits exact self-gravitating solutions that are not simply small deformations of well-known Maxwell-based spacetimes \cite{Bandos:2020jsw}.

A substantial catalog of exact Einstein–ModMax solutions has already been constructed, featuring Reissner–Nordström-type black holes and exact gravitational waves \cite{Flores-Alfonso:2020euz}, Taub–NUT geometries \cite{BallonBordo:2020jtw}, its nonlinear dyonic extensions including NUT wormholes and Taub–Bolt instantons in \cite{Flores-Alfonso:2020nnd}, accelerating black holes \cite{Barrientos:2022bzm}, Melvin–Bonnor/C-metric backgrounds \cite{Barrientos:2024umq}, as well as exact multi–black hole and black dihole configurations \cite{Bokulic:2025usc},  and the Harrison-generated black diholes of \cite{Bokulic:2025ucy}. Concurrently, Ortaggio has recently demonstrated that wide classes of non-null Einstein-Maxwell configurations can be promoted to solutions of conformally invariant nonlinear electrodynamics simply by applying a constant rescaling to the electromagnetic field. This includes, in particular, all static solutions and, more broadly, any configuration that possesses a non-null twistfree Killing vector \cite{Ortaggio:2025sip}. Thus, the search for genuinely nonlinear solution branches without a Maxwell limit is greatly restricted. Yet this space is still dominated by static, accelerating, or NUT-like spacetimes, many closely resembling their Maxwell counterparts. By contrast, the rotating sector is far more rigid in nonlinear electrodynamics: even in the slow-rotation limit, the usual Maxwell-inspired ansatz typically fails for nontrivial nonlinear theories.

This difficulty becomes significantly more severe when both rotation and a dilaton are present. Even within Einstein–Maxwell–dilaton theory, asymptotically flat, charged, rotating black holes are available in closed form only for very specific coupling choices, whereas generic dilatonic couplings have traditionally had to be treated perturbatively and have only recently been realized through fully numerical solutions \cite{Herdeiro:2025blx}. In the nonlinear regime, the rotating sector is known to be much more constrained than the static one: already at the level of slow rotation, the usual Maxwell-type ansatz fails to remain consistent for general nontrivial nonlinear electrodynamics \cite{Lammerzahl:2018zvb}. While recent progress has extended Einstein–ModMax–Dilaton theory to exact extremal charged black holes and black diholes in purely electric or purely magnetic configurations \cite{Bokulic:2025ucy}, an exact rotating dilatonic family is still lacking.

In this work, we employ \emph{the second class of solution} introduced in \cite{Bixano:2026ouq} to construct a exact solution belonging to the nonlinear $\mathcal F/\mathcal G=\mathrm{const}$ sector of Einstein–ModMax-type gravity. This solution exhibits nontrivial electric and magnetic potentials, with both $A_t$ and $A_\varphi$ switched on, and it covers configurations carrying NUT charge as well as a Nut-less, asymptotically flat limit. A key point is that, within our parametrization, this branch cannot be deformed to the Maxwell framework, which shows that it is intrinsically supported by the nonlinear ModMax dynamics rather than being derived from a Maxwell-like branch. Furthermore, in the prolate regime we identify a genuine black-hole sector whose exterior region fulfills the null energy condition, with the curvature singularity remaining hidden behind the event horizon.

In the study \cite{Bixano:2026ouq}, building on the results of \cite{Bixano:2026xum}, the authors developed a framework for obtaining exact solutions in the Einstein–ModMax–scalar field setting, formulated in terms of Ernst-type potentials. In that work, they employ the Lagrangian written in the \textit{Einstein frame}:
\begin{equation}\label{Lagrangiano}
    \mathfrak{L}=\sqrt{-g}\bigg(-\mathcal R +2\epsilon_0 (\nabla \phi)^2 + e^{-2 \alpha_0 \phi } \mathfrak{L}_{MM} \bigg),
\end{equation}
where 
\[
    \mathfrak{L}_{MM}= \mathcal{F} \; \cosh{\gamma} +\sqrt{\mathcal{F}^2+\mathcal{G}^2} \; \sinh{\gamma} .
\]
Here, \(g = \det(g_{\mu\nu})\) is the determinant of the metric tensor, \(\mathcal R\) the Ricci scalar, and \((\nabla \phi)^2 = \nabla_{\mu} \phi \, \nabla^{\mu} \phi\). The symbol \(\gamma\) is the deformation parameter. The Maxwell invariant is \(\mathcal F = F_{\mu\nu} F^{\mu\nu}\), and its dual is \(\mathcal G = F_{\mu\nu} \star F^{\mu\nu}\), with \(\star F^{\mu\nu}\) the dual Faraday tensor. The scalar field is \(\phi\), and the sign parameter \(\epsilon = \pm 1\) distinguishes the dilaton (\(+1\)) from the phantom (\(-1\)) scalar field. The coupling constant \(\alpha_{0} \in \{0, 1, 3, \tfrac{1}{2\sqrt{3}}\}\) selects the theory: Einstein–ModMax (EMM), low-energy effective superstring (SS), Kaluza–Klein (KK), and entanglement relativity (ER), respectively.
The corresponding field equations are:
\begin{subequations}\label{EcuacionesDeCampoOriginales}
\begin{equation}\label{Eq:Campo1}
    \nabla_\mu \left(  P^{\mu \nu} \right)=0,
\end{equation}
\begin{equation}\label{Eq:Campo2}
    \epsilon_0 \nabla^2 \phi+\frac{\alpha_0}{2}  \left( e^{-2\alpha_0 \phi}\, \mathfrak{L}_{MM} \, \right)=0,
\end{equation}
{\small
\begin{equation}\label{Eq:Campo3}
    \mathcal R_{\mu \nu}=2 \epsilon_0 \nabla_\mu \phi \nabla_\nu \phi  + 2 \, w \, e^{-2\alpha_0 \phi} \left( F_{\mu \sigma} \tensor{F}{_\nu}{^\sigma} -\frac{1}{4} g_{\mu \nu } \mathcal{F} \right),
\end{equation}
}
\end{subequations}
where we have defined the following variables to streamline the subsequent calculations
\begin{subequations}
\begin{align}
    &\kappa^2=e^{-2\alpha_0 \, \phi }, \\
    &\mathtt{X} = \kappa^2 \, \mathcal{F}, \qquad  
    \mathtt{Y} = \kappa^2 \, \mathcal{G}, \\ 
    &\Delta =\sqrt{\mathtt{X}^2+\mathtt{Y}^2},\qquad 
    \tan{\Theta}=\frac{\mathtt{Y}}{\mathtt{X}},
\end{align}
\begin{align}
    w(\Theta)&=\cosh{\gamma} +\cos{\Theta} \, \sinh{\gamma} ,
\\
v(\Theta)&=\sin{\Theta} \, \sinh{\gamma} ,
\end{align}
    \begin{equation}
        P^{\mu \nu}=\kappa^2 \Big[ w\, F^{\mu \nu} + v \,  \star F^{\mu \nu}\Big].
    \end{equation}
\end{subequations}
In this notation, the ModMax contribution to the Lagrangian can be written as
\(
    e^{-2 \alpha_0 \phi } \mathfrak{L}_{MM}=w \, \mathtt{X}+v \, \mathtt{Y}.
\)
It is easy to see that switching off the deformation parameter by choosing \(\gamma = 0\) implies \(w = 1\) and \(v = 0\). As a result, we find \(\mathfrak{L}_{MM} = \mathtt{X}\) and \(P^{\mu \nu} = \kappa^2 F^{\mu \nu}\), which precisely reproduces the formulation given in \cite{Bixano:2026xum}.

The framework proposed in \cite{Bixano:2026xum} and \cite{Bixano:2026ouq} relies on an axisymmetric, stationary space-time whose metric is given by
\begin{equation}\label{ds Cilindricas}
     ds^2=-f\left ( dt-\omega d \varphi \right )^2+f^{-1} \left ( e^{2k} (d\rho ^2 + dz^2) +\rho^2 d\varphi^2 \right ), 
\end{equation}
In this formulation, the metric functions $\{f, \omega, \kappa\}$ are assumed to depend solely on the coordinates $(\rho,z)$. We likewise introduce an axisymmetric ansatz for the four-potential, $A_{\mu} = \big[ A_t(\rho,z),\,0,\,0,\,A_{\varphi }(\rho,z) \big]$, which implies that the scalar field depends on the same coordinates, $\phi(\rho,z)$.
%%%%%%%%%%%%%%%%%%%%%%%%%%%%%%%%%%%%%%%%%%%%%%%%%%%%%%%%%%%%%%%%%%%%%%%%%%%%%%%%%%%%%%%%%%%%%%%%%%%%%%%%%%%%%%%%%%%%%%%%%%%%%%%%%%%%
%\section{The new exact solution}
%%%%%%%%%%%%%%%%%%%%%%%%%%%%%%%%%%%%%%%%%%%%%%%%%%%%%%%%%%%%%%%%%%%%%%%%%%%%%%%%%%%%%%%%%%%%%%%%%%%%%%%%%%%%%%%%%%%%%%%%%%%%%%%%%%%%
The new exact solution. 
To construct our new configuration within the branch $\mathcal{F}/\mathcal{G}:$ constant, so that 
\[
    \eta:=\frac{v}{w}=
\frac{\mathtt Y\,\sinh\gamma}
{\sqrt{\mathtt X^2+\mathtt Y^2}\,\cosh\gamma+\mathtt X\,\sinh\gamma}=\eta_0, 
\]remains constant, we make use of the second class of solutions discussed in Ref.~\cite{Bixano:2026ouq}, focusing specifically on the subfamily characterized by nontrivial rotation. The associated solution can be written as:
\begin{subequations}\label{SOLUCION}
    \begin{align}
        f&=
        \frac{f_1}{\mathfrak s_\pm+f_1}, 
        \qquad
        \omega
        =
        \frac{L}{f_1}\,\widetilde{\mathfrak s}_\pm,
        \qquad 
        k=0,
        \\
        A_t&=
        \frac{2}
        {\sqrt{w_0}\,\sqrt{(\sqrt{2}\alpha_0+1)^2+4}}
        \left(
        \frac{f_1}{\mathfrak s_\pm+f_1}
        \right)^{\frac{1+\sqrt{2}\alpha_0}{2}},
        \\
        A_\varphi &= -\frac{\omega}{L} \, A_t, \qquad \phi=-\frac{1}{\sqrt{2}}\ln{ \left( \frac{\mathfrak s_\pm +f_1}{f_1} \right) }
    \end{align}
\end{subequations}
where , for the potentials, we have fixed the gauge by setting $\psi_1=\chi_1=\epsilon_1=0$, and we have chosen $f_1=-s_0$ together with $\kappa_1 = 1/(-s_0)^{\alpha_0/\sqrt{2}}$ so that, in the limit \(x \to \pm \infty\), the function \(f\) tends to one, and the scalar field asymptotically goes to zero. 

In a similar manner, we choose the following harmonic function and its harmonic-dual \footnote{For a more comprehensive discussion, refer to Ref.\cite{Bixano:2026ouq}.} expressed in spheroidal (\(\rho:=L_{\pm}\sqrt{(x^2\pm1)(1-y^2)},\; z:=L_{\pm }x y \; ; \;  L_\pm \in \mathbb{R} \, \& \, L_\pm > 0 \)) oblate ($+$)/prolate ($-$) coordinates \footnote{For a detailed treatment of prolate and oblate spheroidal coordinates and the corresponding $L_\pm$ parameter, refer to \cite{Bixano:2025bio,Bixano:2026ouq,Bixano:2026xum}. }
\begin{subequations}\label{eq:Funciones harmonicas}
    \begin{align}
        \mathfrak s_\pm (x,y)&=\frac{\lambda_0 y+\tau_0 x}{x^2 \pm y^2},
        \\
        \widetilde{\mathfrak s}_\pm (x,y)&=
        \frac{\lambda_0 x(1-y^2)-\tau_0 y(x^2 \pm1)}{x^2 \pm y^2},
    \end{align}
\end{subequations}
where the metric \eqref{ds Cilindricas}, takes the form:
\begin{multline}\label{ds sp}
    ds^2 = -f\left( dt-\omega d \varphi \right)^2
     + \frac{(L_{\pm})^2}{f} \bigg( (x^2\pm1)(1-y^2) d\varphi^2 \\
     +(x^2\pm y^2) e^{2k} \left\{ \frac{dx^2}{x^2\pm 1} +\frac{dy^2}{1-y^2} \right\} \bigg).
\end{multline}

The exact solution \eqref{SOLUCION} exists only for $\epsilon_0=+1$, that is, in the dilatonic sector. By inserting \eqref{SOLUCION} and \eqref{eq:Funciones harmonicas} into \eqref{ds sp}, and then substituting the result into \eqref{EcuacionesDeCampoOriginales} expressed in spheroidal coordinates, the full system of field equations is fulfilled if and only if
\begin{equation}\label{RESTRICCION}
    \eta_0=\frac{2\alpha_0^2-5}{2\sqrt{2}\,\alpha_0}.
\end{equation}
Hence, for this exact solution, the ModMax parameter $\eta_0$ is not free, but is fixed by the dilatonic coupling $\alpha_0$ of the theory.

The Maxwell framework is characterized by $\eta_0=0$, which, by \eqref{RESTRICCION}, is reached only at the isolated coupling
\begin{equation}
    \alpha_0{}^2=\frac52.
\end{equation}
\emph{Thus, within this specific family, the solution is generally non-Maxwellian: for any dilatonic coupling, including, in particular, the usual low-energy string and Kaluza-Klein values, it inevitably belongs to the nonlinear ModMax regime. Consequently, the configuration should be viewed as inherently ModMax, whereas a Maxwell description, if it occurs at all, is confined to the isolated point $\alpha_0^2 = 5/2$.}
%------------------------------------------------------------
%\subsection{Singularities}
%------------------------------------------------------------
Singularities. The Ricci invariant can be expressed by the following formula
\begin{widetext}
\begin{align}
    \mathcal{R}=\frac{
    4\lambda_0\tau_0xy(x^2\mp y^2)
    +\tau_0^2\bigl[x^4+y^2\pm x^2(1-3y^2)\bigr]
    +\lambda_0^2\bigl[x^2(1+3y^2)\pm y^2(1-y^2)\bigr]
    }{
    L_\pm^2
    (x^2\pm y^2)
    \bigl(\tau_0x+x^2+y(\lambda_0\pm y)\bigr)^3
    },
\end{align}
\end{widetext}
and the associated Kretschmann invariant can be written as
{\footnotesize
\[
    \mathcal{K}=\mathcal{R}_{\mu \nu \alpha \beta}\mathcal{R}^{\mu \nu \alpha \beta} \propto \frac{\mathcal{N}(x,y)
    }{
    L_\pm^4
    (x^2\pm y^2)^2
    \bigl(\tau_0x+x^2+y(\lambda_0\pm y)\bigr)^6
    },
\]
}
where the denominator has a higher overall power than the numerator \(\mathcal{N}(x,y)\).

Hence, there is a ring singularity at the point $x=0$, $y=0$, along with two surface singularities given by $x=\pm y$ in the prolate branch. These latter two have been studied in greater detail in \cite{Bixano:2025bio,Bixano:2025qxp}. In the present case, an additional singularity arises exactly when \(\tau_0x + x^2 + y(\lambda_0 \pm y) = 0\). For the lower-sign branch,
\(
x^2+\tau_0x+\lambda_0y-y^2,
\)
and for \(x\ge1\), \(y\in[-1,1]\), one has
{\small
\begin{align*}
    x^2+\tau_0x+\lambda_0y-y^2
    & \ge 
    x^2+\tau_0x-\lambda_0-1 \\
    &=(x-1)(x+1+\tau_0)+(\tau_0-\lambda_0).
\end{align*}
}
Consequently, if \(\lambda_0<\tau_0\), this expression is strictly positive for every \(x\ge1\), and thus the roots
\[
x_{\pm}^{(-)}
=
\frac{-\tau_0 \pm \sqrt{\tau_0^2 - 4y(\lambda_0 - y)}}{2}
\]
do not lie in the physical region. In the limiting case \(\lambda_0=\tau_0\), the expression is merely non-negative and becomes zero only at
\(
x=1,
\qquad
y=-1,
\)
i.e. exactly at the boundary pole. In contrast, for the branch with the plus sign,
\(
x^2 + \tau_0 x + \lambda_0 y + y^2,
\)
the quantity stays strictly positive for all \(x\ge1\) under the same conditions, so no roots appear there, even when \(\lambda_0=\tau_0\). In other words, the third surface singularity exists within the region $x < 1$, whenever \(\lambda_0<\tau_0\) .
%------------------------------------------------------------
%\subsection{Conserved charges}
%------------------------------------------------------------
Conserved charges. To compute the conserved charges, we will explicitly follow the method introduced in \cite{Bixano:2025bio}, Section V and Appendix A. Here we adopt the definitions
{\small
\begin{align*}
    &M(x):=-\frac{1}{8\pi}\int_{S_x}{}^\star d\xi^\flat,
    \qquad
    N(x):=\frac{1}{8\pi}\int_{S_x}d\xi^\flat,
    \\
    &J(x):=\frac{1}{16\pi}\int_{S_x}{}^\star d\varsigma^\flat,
    \qquad
    H(x):=\frac{1}{4\pi}\int_{S_x}F,
    \\
    &Q^{\mathrm{EMD}}(x):=\frac{1}{4\pi}\int_{S_x}\kappa^2\,{}^\star F,
    \qquad
    Q^{\mathrm{MM}}(x):=
\frac{1}{4\pi}\int_{S_x}{}^\star P,
\end{align*}
}
where $EMD$ denotes Einstein-Maxwell-Dilaton, $MM$ detones ModMax, the two-dimensional surface is given by
\(
S_x:\quad t=\text{const},\quad x=\text{const},\quad (y,\varphi)\in[-1,1]\times[0,2\pi),
\)
and $\xi$ denotes the one-form corresponding to the Killing vector \(\partial_t\), while $\varsigma$ denotes the one-form corresponding to the Killing vector \(\partial_\varphi\). In this way, we arrive at the following conserved charges, expressed in terms of \(\mathtt F_\infty=\lim_{x\rightarrow\pm \infty}\mathtt F (x)\), where $\mathtt F\in \{ M,N,J,Q^{EMD},Q^{MM} \} $ are
{\small
\begin{subequations}\label{eq:CargasInvariantes}
    \begin{equation}
    N_\infty=-\frac{L_\pm\tau_0}{2f_1},
    \qquad
    M_\infty=\frac{L_\pm\tau_0}{2f_1},
    \qquad
    J_\infty=-\frac{L_\pm^2\lambda_0}{2f_1},
    \end{equation}
    \begin{equation}
    H_\infty=
    \frac{2L_\pm}{\sqrt{w_0}\sqrt{(\sqrt{2}\alpha_0+1)^2+4}}
    \frac{\tau_0}{f_1},
    \end{equation}
    \begin{equation}
    Q_\infty^{\mathrm{EMD}}
    =
    \frac{(\sqrt{2}\alpha_0+1)L_\pm}{\sqrt{w_0}\sqrt{(\sqrt{2}\alpha_0+1)^2+4}}
    \frac{\tau_0}{f_1},
    \end{equation}
    \begin{equation}
    Q_\infty^{\mathrm{MM}}
    =
    \frac{L_\pm\sqrt{w_0}}{2\sqrt{2}\alpha_0}
    \sqrt{(\sqrt{2}\alpha_0+1)^2+4}\,
    \frac{\tau_0}{f_1}.
    \end{equation}
\end{subequations}
}
%------------------------------------------------------------
%\subsection{Event horizon in the prolate configuration}
%------------------------------------------------------------
Event horizon in the prolate configuration. In the prolate sub-extreme regime, an event horizon is present and can be straightforwardly determined by following the procedure outlined in Ref. \cite{Bixano:2025bio,Bixano:2025qxp}. This horizon is obtained specifically in the limit \(\rho=0\). Therefore the surface \(x=1\) naturally emerges as the candidate horizon. In addition, for the specific solution considered here, one finds \(\widetilde{\mathfrak s}(1,y)=\lambda_0\), which does not depend on \(y\). Consequently, the dragging function evaluated on the horizon,
\begin{equation}
\omega_H=\omega(1,y)=\frac{L_\pm\lambda_0}{f_1},
\end{equation}
is constant over the whole surface \(x=1\). Therefore, the Killing vector
\begin{equation}
\chi_H=\partial_t+\Omega_H\partial_\varphi,
\qquad
\Omega_H=\frac{1}{\omega_H}=\frac{f_1}{L_\pm\lambda_0},
\end{equation}
becomes null on \(x=1\), indicating that the prolate branch indeed possesses an event horizon located at \(x_H=1\).In Boyer–Lindquist coordinates (\(L_\pm x = r - l_1 \, ; \, y=\cos{\theta}\)), and making use of the relation \(L_\pm x = r - l_1\), the event horizon (located at \(x_H = 1\)) isequivalent to
\begin{equation}
r_H=l_1+L_\pm.
\end{equation}
%------------------------------------------------------------
%\subsection{Null Energy Condition}
%------------------------------------------------------------
Null Energy Condition. To derive the null energy condition, we follow the framework and techniques outlined in \cite{Bixano:2025bio}, Section VII, where the null energy condition (\(T_{\mu \nu} l^{\mu} l^{\nu} \geq 0 \; ; \; l^{\mu}l_{\mu}=0 \) ) is derived.
{\small
\begin{widetext}
\begin{align}\label{eq:NEC}
T_{\mu \nu} l^{\mu} l^{\nu}
=&
\frac{
4\lambda_0\tau_0xy
\bigl[
x^4-y^4+(y^2\mp x^2)(1-y^2)
\bigr]
+2\lambda_0^2
\bigl[
x^4(1+y^2)+y^4(1-y^2\pm 2x^2)
\bigr]
}{
L_\pm^2
(x^2\pm y^2)^2
\bigl(
\tau_0x+x^2+y(\lambda_0\pm y)
\bigr)^3
} \nonumber
\\
&+\frac{\tau_0^2
\bigl[
x^6
\pm x^4(1-2y^2)
+6x^2y^2(1-y^2)
\pm y^4(1\mp x^2)
\bigr]}{
L_\pm^2
(x^2\pm y^2)^2
\bigl(
\tau_0x+x^2+y(\lambda_0\pm y)
\bigr)^3
}.
\end{align}
\end{widetext}
}
It is also useful to examine the numerator. In the domain
\(
x\ge1,
\quad
y\in[-1,1],
\)
all three bracketed factors are non-negative for both the upper-sign and lower-sign cases. In particular, for the mixed term we have
\(
x^4-y^4+(y^2\mp x^2)(1-y^2) \geq0
\)
and is non-negative for \(x\ge1\), \(|y|\le1\). For the \(\lambda_0^2\)-term,
\(
x^4(1+y^2)+y^4(1-y^2+2x^2)\ge0
\)
for the upper sign, while for the lower sign
\(
x^4(1+y^2)+y^4(1-y^2-2x^2)\ge0,
\)
since, viewed as a quadratic polynomial in \(x^2\), its minimum over \(x^2\ge1\) is attained at \(x=1\), where it becomes
\(
1+y^2-y^4-y^6=(1-y^2)(1+y^2)^2\ge0.
\)
Finally, let \(u=y^2\), so that \(0\le u\le1\). With this substitution, the \(\tau_0^2\)-term becomes a quadratic polynomial in \(u\). For the plus sign, we obtain
\(
\mathfrak Q_+(u)=x^6+x^4+(6x^2-2x^4)u+(1-7x^2)u^2,
\)
while for the minus sign, we have
\(
\mathfrak Q_-(u)=x^6-x^4+(2x^4+6x^2)u-(1+7x^2)u^2.
\)
Since \(x\ge1\), the coefficient of \(u^2\) is negative in both cases, which means each quadratic is concave on the interval \(0\le u\le1\). Therefore, the minimum value on this interval must occur at one of the endpoints. Evaluating at these endpoints yields
{\small
\[
\mathfrak Q_+(0)=x^4(x^2+1)\ge0,
\qquad
\mathfrak Q_+(1)=(x^2-1)^2(x^2+1)\ge0,
\] }
and 
{\small
\[
\mathfrak Q_-(0)=x^4(x^2-1)\ge0,
\qquad
\mathfrak Q_-(1)=(x^2-1)(x^2+1)^2\ge0.
\]
}
Therefore, the \(\tau_0^2\) term is non-negative for either sign over the entire domain \(x \ge 1\), \(y \in [-1,1]\).

For the negative-sign choice in \eqref{eq:NEC}, corresponding to the prolate configuration (lower sign), we now assume
\(
0<\lambda_0\le\tau_0<1,
\quad
x\ge1,
\quad
y\in[-1,1].
\)
Then the denominator is non-negative throughout the whole domain, since
\(
L_\pm{}^2  (x^2\pm y^2)^2\ge0,
\)
and
{\small
\begin{align*}
x^2+\tau_0x+\lambda_0 y-y^2
&\ge
x^2+\tau_0x-\lambda_0-1 \nonumber\\
&=
(x-1)(x+1+\tau_0)+(\tau_0-\lambda_0)
\ge0.
\end{align*}
}
Thus, the denominator is strictly positive, and consequently \(T_{\mu \nu} l^{\mu} l^{\nu} \geq 0\) holds throughout the entire domain
\(
0<\lambda_0\le\tau_0<1,
\quad
x\ge1,
\quad
y\in[-1,1]
\).

%%%%%%%%%%%%%%%%%%%%%%%%%%%%%
%\section{Conclusions}   %
%%%%%%%%%%%%%%%%%%%%%%%%%%%%%
Conclusions. We have introduced a new exact solution of the system \eqref{EcuacionesDeCampoOriginales} for the Einstein–ModMax–Dilaton framework, featuring non‑zero rotation ($\omega\neq0$) and a dilatonic scalar field ($\phi \neq 0$), and, more importantly, belonging to the branch with \(\mathfrak F/ \mathfrak G:\) constant. A key aspect of this solution is that the ModMax parameter (\(\eta_0\)), which encodes the deformation of Maxwell theory, cannot vanish, see eq. \eqref{RESTRICCION}. In other words, this configuration does not allow the Einstein–Maxwell–Dilaton limit (\(\eta_0 = 0\)), except if one assumes an as‑yet unknown theory with coupling \(\alpha_0{}^2 = 5/2\). Consequently, for established models such as Kaluza–Klein \(\alpha_0{}^2 = 3\), low‑energy superstring theory \(\alpha_0{}^2 = 1\), and other familiar cases, the parameter \(\eta_0\) must always remain non‑zero.

Although no corresponding solution exists within the Einstein–Maxwell–Dilaton framework, the ModMax configuration under consideration exhibits three curvature singularities in the oblate branch \((+)\) and five curvature singularities in the prolate branch \((-)\). In the prolate branch, an event horizon is present, and thus the configuration can be identified as a exact rotating dilatonic black hole in the ModMax theory. 

Both the oblate and prolate branches satisfy the null energy condition (NEC) provided that the parameters obey \(0 < \lambda_0 \le \tau_0 < 1\). In particular, for the black-hole solution, this compact object respects Penrose’s weak cosmic censorship conjecture (CCC): all singularities are contained within the event horizon, again under the condition \(0 < \lambda_0 \le \tau_0 < 1\). In the limiting case \(\lambda_0 = \tau_0\), the NEC is violated at the poles, and some singularities likewise intersect the event horizon at the poles.

Finally, this solution exhibits a particularly noteworthy feature: by analyzing the invariant charges \eqref{eq:CargasInvariantes}, we observe that the NUT parameter, the Komar mass, the Einstein–Maxwell–Dilaton (EMD)–type electric charge, and the ModMax magnetic and electric charges are all controlled by the same parameter \(\tau_0\). Consequently, setting \(\tau_0 = 0\) yields an asymptotically flat spacetime and simultaneously forces all of these quantities to vanish, leaving only the Komar angular momentum \(J_\infty\) nonzero. In this regime, the configuration describes a pure mass dipole accompanied by ModMax electric and magnetic dipole moments. Thus, even after turning off \(\tau_0\), the solution remains nontrivial, despite the vanishing of the Komar mass, the EMD electric charge, and the Maxwell and ModMax electric charges. Furthermore, an important observation is that the EMD charge and the ModMax electric charge cannot be made equal, which provides a strong indication that an analogous solution does not exist in the EMD framework.

Finally, this family exhibits an unusually clean separation between its monopolar and rotational sectors. As can be read off from the invariant charges \eqref{eq:CargasInvariantes}, the NUT parameter, the Komar mass, the magnetic flux charge, the EMD-like electric charge, and the ModMax electric charge are all proportional to a single parameter \(\tau_0\), while the Komar angular momentum \(J_\infty\) is independently determined by \(\lambda_0\). Hence, \(\tau_0\) controls the complete monopolar charge content of the configuration, whereas \(\lambda_0\) governs its genuinely rotational degrees of freedom. In particular, imposing \(\tau_0=0\) simultaneously eliminates the NUT charge, the Komar mass, the magnetic charge, the EMD electric charge, and the ModMax electric charge, yet leaves \(J_\infty\neq0\). The ensuing geometry is therefore asymptotically flat and still nontrivial: the entire monopolar sector may be switched off without erasing the rotational structure of the solution. This strict decoupling between monopolar and rotational data is itself highly non-generic and of considerable structural significance.

%A second decisive point is that the EMD electric charge and the ModMax electric charge do not coincide in general. Therefore, the present family cannot be interpreted as a simple Einstein-Maxwell-Dilaton configuration rewritten in different variables. Rather, it defines a genuinely ModMax-supported branch, with asymptotic charge data that are intrinsically distinct from their EMD counterparts. Moreover, in the prolate sector, the condition \(\tau_0\ge \lambda_0\) identifies the black-hole regime compatible with both positivity of the NEC in the exterior region and cosmic censorship, while the saturated case \(\tau_0=\lambda_0\) is critical, in the sense that the curvature singular set \(\Sigma=0\) only touches the horizon at its boundary. Altogether, these results show that the family constructed here is not only exact and nontrivial, but also physically structured in a highly constrained and interpretable way: the parameters separately control charge, rotation, and horizon regularity, and the ModMax sector introduces effects that are not reproducible within the standard EMD framework.

%------------------------------------------------------------
%\vspace{-6pt}
%\section{Acknowledgements}

Acknowledgements. LB thanks SECIHTI-M\'exico for the doctoral grant.
This work was also partially supported by SECIHTI M\'exico under grants SECIHTI CBF-2025-G-1720 and CBF-2025-G-176. 

%The authors are gratefully for the computing time granted by LANCAD and CONACYT in the Supercomputer Hybrid Cluster "Xiuhcoatl" at GENERAL COORDINATION OF INFORMATION AND COMMUNICATIONS TECHNOLOGIES (CGSTIC) of CINVESTAV. URL: http://clusterhibrido.cinvestav.mx/ and to Hector Oliver Hernandez for his help with the code installations.

%%%%%%%%%%%%%%%%%%%%%%%%%%%%%%%%%%%%%%%%%%
%\begin{appendices}
%%%%%%%%%%%%%%%%%%%%%%%%%%%%%%%%%%%%%%%%%%%%%%%%%%%%%%%%%%%%%%%%%%%%%%%%%%%%%%%%%%%%%%%%%%%%%%%%%%%%%%%%%%%%%%%%%%%%%%%%%%%%%%%%%%%%%%%%%%%%%%%%%%%%%%%%%%%%%%%%%%%%%%%%
%\section{Coordinates protection}\label{Apendix:Coordinates protection}
%%%%%%%%%%%%%%%%%%%%%%%%%%%%%%%%%%%%%%%%%%%%%%%%%%%%%%%%%%%%%%%%%%%%%%%%%%%%%%%%%%%%%%%%%%%%%%%%%%%%%%%%%%%%%%%%%%%%%%%%%%%%%%%%%%%%%%%%%%%%%%%%%%%%%%%%%%%%%%%%%%%%%%%%
 
%\end{appendices}
%%%%%%%%%%%%%%%%%%%%%%%%%%%%%%%%%%%%%%%%%%
\bibliographystyle{elsarticle-harv} 
\bibliography{Bibliografia}

\end{document}